\documentclass[aip,cha,author-year,preprint,onecolumn,groupedaddress,amsmath,amssymb,floatfix]{revtex4-1}
\usepackage{graphicx}
\usepackage{graphics}

\begin{document}


\title{Viscoelasticity of a colloidal gel during dynamical arrest: evolution through the critical gel and comparison with a soft colloidal glass}


\author{Ajay Singh Negi}
\email[]{ajaynegi@iiserb.ac.in}
\affiliation{Department of Physics, Indian Institute of Science Education and Research, Bhopal India}
\author{Carissa G. Redmon}
\email[]{carissa1.redmon@famu.edu}
\affiliation{Department of Chemical and Biomedical Engineering, FAMU-FSU College of Engineering, Tallahassee FL, 32310 USA}
\author{Subramanian Ramakrishnan}
\email[]{sramakrishnan@fsu.edu}
\affiliation{Department of Chemical and Biomedical Engineering, FAMU-FSU College of Engineering, Tallahassee FL, 32310 USA}
\author{Chinedum O. Osuji}
\email[]{chinedum.osuji@yale.edu}
\affiliation{Department of Chemical Engineering, Yale University,New Haven CT 06511}
\author{}
\affiliation{}


\date{\today}

\begin{abstract}
We consider the gelation of colloidal particles in suspension after cessation of shear flow. Particle aggregation is driven by a temperature-tunable attractive potential which controls the growth of clusters under isothermal conditions. A series of frequency resolved time sweeps is used to systematically reconstruct the frequency dependent dynamic moduli as a function of time and temperature or attraction strength. The data display typical hallmarks of gelation with an abrupt transition from a fluid state into a dynamically arrested gel state after a characteristic gelation time $t_g$ that varies exponentially with temperature and serves to collapse the evolution of the system onto a universal curve. We observe the viscoelastic properties of the critical gel where we find that $G'(\omega)\approxeq G''(\omega)\sim \omega^{n_c}$ where $n_c=0.5$ in a narrow time window across all attraction strengths. We measure a dynamic critical exponent of $\kappa=0.25$ which is similar to that observed in crosslinked polymer gels. The approach to the critical gel is therefore governed by $\eta_0\sim-\epsilon^{-s}$ and $G_e\sim\epsilon^z$ with $s=z=2$ where $\epsilon=p/p_c-1$ is the distance to the gel point. Remarkably, the relaxation moduli of the near-critical gels are identical across the temperatures considered, with $G(t)\approx0.33t^{-0.5}$. This suggests an underlying strong similarity in gel structure in the regime of attraction strengths considered, despite the differences in aggregation kinetics. We contrast these findings with the behavior of a colloidal glass undergoing dynamical arrest where no critical state is observed and where the arrest time of the system displays a marked frequency dependence. These findings highlight the underlying structural differences between colloidal gels and glasses which are manifest in their dynamic properties in the vicinity of the liquid-to-solid transition.
\end{abstract}


\maketitle


\section{\label{introduction}Introduction}

The aggregation of colloidal particles experiencing short-range attractive interactions may lead to the emergence of arrested states at low volume fractions due to the eventual formation of space spanning structures.[\cite{zaccarelli2007colloidal}] These dilute colloidal gels exhibit a tenuous network-like morphology with fractal dimensions that can be related to the conditions of the aggregation, i.e. diffusion- or reaction-limited aggregation, DLCA and RLCA respectively. Pure DLCA leads to fractal dimension $d_f\approxeq1.8$ [\cite{Weitz1984fractal,Weitz_DLCA1989}] and a monodisperse cluster size distribution while pure RLCA leads to more compact and size-disperse clusters with $d_f\approxeq2.1$.[\cite{Meakin1988}] In dense suspensions, hard sphere excluded volume interactions as well as variable-ranged attractive or repulsive interactions support arrested states with a liquid-like structure factor, colloidal glasses.[\cite{Pusey_Review2008,HunterWeeks2012}] In both cases, starting from an initially ergodic system of dispersed particles, there is a transition over time to a dynamically arrested or non-ergodic state due to the localization of particles by attractive interactions, bonding, with other particles, or caging by neighboring particles. We may speak in broad terms of a transition from a liquid-like to a solid-like state coinciding with the drastic slowing of the system dynamics on entry into the arrested glass or gel. The transition from a dilute colloidal gel to an attractive glass at constant attraction strength is continuous and subtle in nature. Reflecting this, the rheological properties of colloidal gels at intermediate volume fractions, \textit{ca}. $0.20<\varphi<0.45$, exhibit complexity that reflects the role of bonding both among and within particle clusters in dictating the dynamical and mechanical properties. One consequence of this for example is the emergence of 2-step yielding in the nonlinear rheology.[\cite{koumakis2011two}] While the distinctions between colloidal gels at intermediate volume fraction and attractive glasses are not always sharp, it is clear that the relevance of the ramified network picture in describing the structure of attractively interacting systems diminishes progressively with increasing volume fraction.[\cite{laurati2011nonlinear}]

There is considerable interest in understanding the nature of the liquid-to-solid transitions that occur in colloidal systems as a function of their composition and the inter-particle interactions. This interest is motivated in part by the heavy utilization of colloidal dispersions in consumer and industrial products [\cite{mezzenga2005understanding}] as well as by the intriguing similarities between the physics of colloidal systems and atomic and molecular counterparts.[\cite{anderson2002insights,HunterWeeks2012}] It is useful to develop a physical picture for the state of the system at the precise moment that it transitions from an liquid-like material to a non-ergodic soft solid for an attractive system, i.e. the gel point. Multiple frameworks have been proposed for gelation, ranging from equilibrium situations of particle aggregation in the absence of phase separation leading to the formation of a system-spanning cluster, to non-equilibrium scenarios including jamming of clusters [\cite{trappe2001jamming}] and spinodal decomposition at higher volume fractions arrested by a jamming transition within a colloid-rich phase.[\cite{Manley2005glasslike,Cardinaux2007,lu2008gelation}] These scenarios have been comprehensively reviewed [\cite{zaccarelli2007colloidal}] but it remains the case that clear consensus on this topic has yet to emerge. These differences notwithstanding, at the gel point there exist tenuous physical connections among particles over a length scale that is sufficient to span the system and to transmit stress across it, thus conferring elasticity. In a formal sense, gelation is a critical phase transition in connectivity and this intermediate state right at the gel point, neither liquid nor solid, is called the critical gel.[\cite{Winter1997Adv}]

\begin{eqnarray}
\text{Sol, $p<p_c$}&\begin{cases}
\eta_0&\sim(p_c-p)^{-k}\\
\lambda_{max}&\sim(p_c-p)^{-\alpha_-}\\
\label{eq:diverging_viscosity}
\end{cases}\\
\text{Gel, $p>p_c$}&\begin{cases}
G_e&\sim(p-p_c)^{z}\\
\lambda_{max}&\sim(p-p_c)^{-\alpha_+}
\end{cases}
\label{eq:diverging_modulus}
\end{eqnarray}

Percolation theory provides scaling laws for the evolution of static and dynamic properties in the vicinity of the gel point, as a function of the reaction coordinate or extent of reaction $p$ and the normalized distance from the gel point, $\epsilon=p/p_c-1$.[\cite{Stauffer1982}] These relations have been developed strictly speaking for the case of a chemical gel where the permanence of bonds imposes coincidence between the formation of a physically connected network and the liquid to solid transition or dynamical arrest of the system. For physical gels, the finite bond lifetime between constituents represents a complication, but for sufficiently long-lived associations relative to the experimental measurement window, the scaling descriptions maintain their relevance. The zero shear viscosity $\eta_0$ leading up to the critical state diverges as a power-law with $\epsilon$, Equation \ref{eq:diverging_viscosity}. Likewise, the equilibrium modulus $G_e$ also exhibits power-law divergence away from the critical state, Equation \ref{eq:diverging_modulus}. The longest relaxation time $\lambda_{max}$ diverges approaching the gel point as connectivity increases and the largest cluster grows. Post-gelation, the longest finite relaxation time decreases as additional elements are incorporated into the network which has an infinite relaxation time. At the critical point, the gel is a self-similar structure and displays power-law behavior with $G^*(\omega)\sim\omega^{n_c}$.

In the near critical state the complex modulus is assumed to scale as a power law with a cut-off function that depends on the characteristic time $t^*$ or frequency $\omega^*=1/t^*$, $G^*(\omega)=\omega^{n_c}f^{\pm}(\omega/\omega^*)$, with $f^-$ and $f^+$ referring to the pre- and post- critical states. The critical slowing down of the system represents a divergence of this characteristic time, $t^*\sim\epsilon^{-1/\kappa}$. The frequency dependent growth rate of the complex modulus of the critical gel reflects the evolution of the characteristic time through the dynamic critical exponent, $\kappa$, $\left(\partial \log{G^*}/\partial t\right)_{t_g}\sim \omega^{-\kappa}$. The critical scaling exponents $k$, $n_c$ and $z$ can be related to $\kappa$ through an assumption of symmetric divergence of the characteristic time on either side of the gel line ($\epsilon=0$) and imposing the constraint that the viscosity is well-defined for $\omega\rightarrow 0$ and the shear modulus above the critical point remains finite.[\cite{durand1987frequency,Scanlan1991}] These constraints require respectively that $f^-\sim\omega^{1-n_c}$ and $f^+\sim\omega^{-n_c}$. In turn, this implies that $\eta\sim(t^*)^{1-n_c}$ and $G_e\sim(t^*)^{n_c}$. Thus based on the scaling assumption for the dynamic modulus we recover 2 relationships connecting the 4 fundamental scaling exponents for the critical gel, Equation \ref{eq:exponents}, so only 2 are independent. Application of the above scaling arguments to microscopic models for the system dynamics produces numerical values for the critical exponents. For example, the Rouse model predicts $k=4/3$ and $z=8/3$, or $n_c=2/3$ and $\kappa=1/4$.[\cite{martin1989viscoelasticity}] In practice observed values vary over quite a range.[\cite{Winter1997Adv}]

\begin{eqnarray}
k&=&(1-n_c)/\kappa \nonumber \\
z&=&n_c/\kappa
\label{eq:exponents}
\end{eqnarray}

The critical gel has a fragile, fleeting nature.  This makes accurate determinations of the gel point, $p_c$ or gel time, $t_g$ based on observations of power-law divergence of $\eta_0$ or $G_e$ quite challenging. This is particularly the case in physical gels where the weak bonds between structures can be easily disrupted by rheological measurements. Determinations of $G^*(\omega)$ are more viable for both chemical and physical gels given the possibility to work at very small deformations, although the small yield stresses of near critical states still present challenges for physical gels. A once-common approach involves denoting the time at which the elastic modulus $G'$ exceeds the viscous modulus $G''$, i.e. the cross-over time, $t_{cross}$, as the gel time. The cross-over time however is often empirically observed to be a function of the measurement frequency. This is incongruent with the simple fact that there is a well-defined time at which the physical state discussed above must exist, i.e. $t_g$ cannot be a function of $\omega$.

Proper identification of $t_g$ is made possible using the rigorous treatment provided by Winter and Chambon. As mentioned, the critical gel bears an important distinguishing feature in that it displays self-similar structure and dynamics. This is manifest in power law relaxation, with the relaxation modulus $G(t)=S_c t^{-n_c}$ where $n_c$ is the critical relaxation exponent and $S_c$ in $\mathrm{Pa.s^{-n}}$ is termed the critical gel strength. Typically it is observed that $0<n_c<1$ and $n_c\sim 1/S_c$. So called ``strong'' critical gels have $n_c\rightarrow 0$ and weak gels have $n_c\rightarrow 1$.[\cite{Winter1997Adv}] The gel strength is direct measure of the relaxation modulus of the system at $t=1$s and as such it is related to the concentration or the inverse mobility of the system. A constitutive equation for the stress can be derived using Boltzmann linear superposition, Equation \ref{eq:Boltzmann_LVE}.

\begin{equation}
\tau(t)=\int^t_{-\infty}\,dt'S_c(t-t')^{-n_c}\,\dot\gamma(t')
\label{eq:Boltzmann_LVE}
\end{equation}

The dynamic mechanical properties are then as given below

\begin{eqnarray}
G'(\omega)&=&\Gamma(1-n_c)S_c\,\omega^{n_c}\cos(n_c\frac{\pi}{2}) \nonumber \\
G''(\omega)&=&\Gamma(1-n_c)S_c\,\omega^{n_c}\sin(n_c\frac{\pi}{2}) \nonumber \\
\delta&=&n_c\frac{\pi}{2} \nonumber \\
\mid G^*(\omega)\mid &=&\Gamma(1-n_c)S_c\,\omega^{n_c}
\label{eq:dynamic_moduli_critical_gel}
\end{eqnarray}

where the gamma function can be approximated as in Equation \ref{eq:gamma_function} for $0<n_c<1$.

\begin{equation}
\Gamma(n_c)=\frac{1}{n_c}\Gamma(1+n_c)\approx\frac{1}{n_c}\left[1-0.1138\left(1-4(n_c-0.5)^2\right)\right]
\label{eq:gamma_function}
\end{equation}

\begin{figure}
\includegraphics[width=60mm, scale=1]{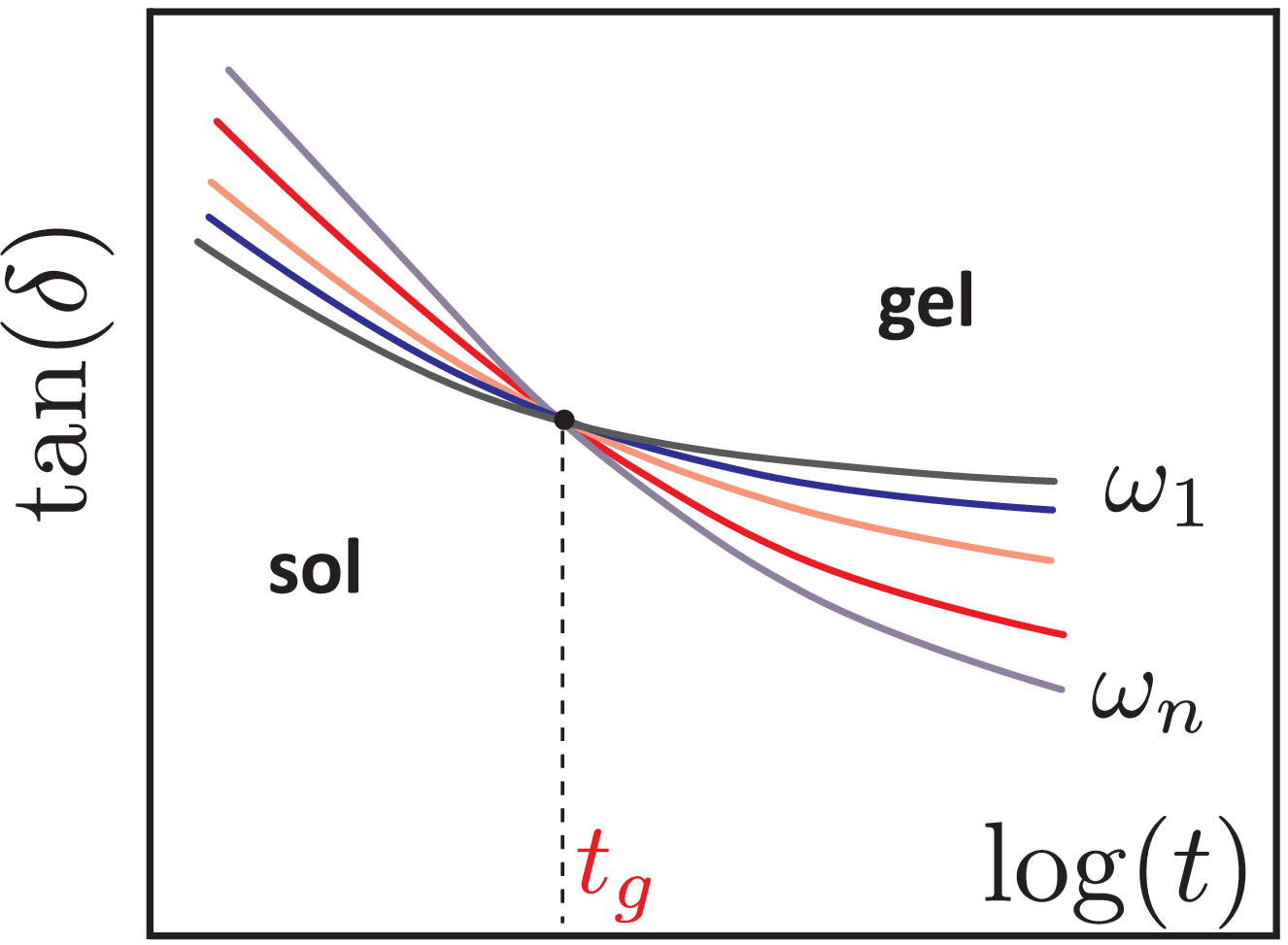}
\includegraphics[width=60mm, scale=1]{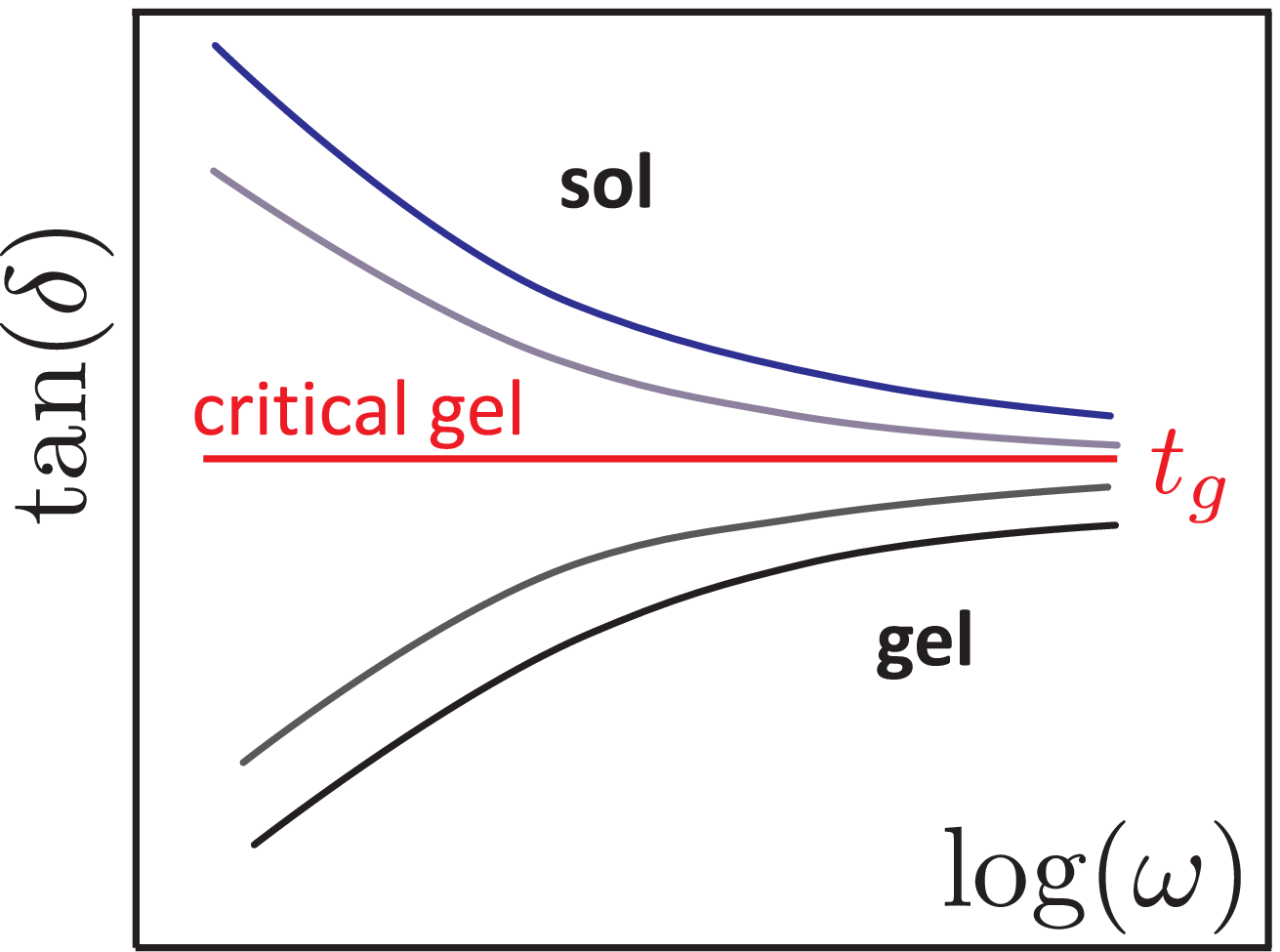}
\caption{Identification of the critical gel and $t_g$ using frequency and time resolved measurements}
\label{tan_delta_gelation_theory}
\end{figure}

In this framework, the cross-over time of the normalized shear moduli $G'/[\cos(n_c\pi/2)]$ and $G''/[\sin(n_c\pi/2)]$ becomes frequency independent and provides the location of $t_g$. Identification of the gel point can be made by measurement of the frequency dependent moduli, with the incipient gel displaying $G'(\omega)\approx G''(\omega)\sim\omega^{n_c}$ at $t=t_g$ with $0<n<1$. $t_g$ is located by the intersection of data in $\tan\delta(t)$ measured at different frequencies or equivalently, by the timepoint at which $\tan\delta\sim\omega^0$, Figure \ref{tan_delta_gelation_theory}. Time-resolved mechanical spectroscopy has been widely applied in polymer systems undergoing chemical gelation, relying in some cases on multiwave excitation to enable data collection across multiple frequencies in a single time sweep measurement. A large body of literature exists describing the scaling behavior of dynamic properties of chemical gels of cross-linked polymers where $n_c$ has been typically observed to be $0.5\leq n_c\leq 1$ and commonly $n_c\approx 0.7$.[\cite{Winter1997Adv}] Comparable studies have been advanced for physical gels with robust interaction strengths such as in semi-crystalline polymers [\cite{Lin1991,Aoki_Li1997a}],  bio-polymers [\cite{Ross-Murphy1991}] and block copolymer or micellar gels. [\cite{Seitz2007,He_Lodge2007}] By contrast, the evolution of viscoelasticity during gelation and the properties of the critical state for particulate gels, where interactions tend to be much weaker, has remained relatively unexplored. In such a system where the gel line is sought as a function of volume fraction, $\varphi_g$, temperature $T_g$ or attraction strength $U_g$, for example, one can make careful measurements of $G^*(\omega)$ across a broad range of nominally equilibrated samples [\cite{grant1993volume,rueb1997viscoelastic,Rueb_Zukoski1998,Cao2010,KrishnaReddy2012}]. To the extent that the systems in question are well-equilibrated and do not display time-dependent behavior, this constitutes a robust approach. Rueb and Zukoski examined dense suspensions $(\varphi=0.447$) of octadecyl coated silica particles in decalin and tetradecane and found $n_c\approx 0.75$ at the gel point. A rheological study by Khan \textit{et al}. of thermoreversible gels of poly(ethylene oxide) coated polystyrene spheres with $0.03\leq\varphi\leq0.27$ uses this approach to identify $T_g$ [\cite{Shay2001}]. Recent work by Vermant \textit{et al}. considered the relaxation exponent $n$ as well as the gel strength $S$ for near-critical thermogels formed by aggregation of polymer-coated rodlike  particles of \textit{fd}-virus.[\cite{KrishnaReddy2012}] They investigated suspensions from 1-8\% w./v. and observed $n\sim S^{-1}$, as expected, with $0.10\leq n\leq 0.35$. The inverse relationship between $n$ and $S$ has been documented in end-linked PDMS polymer gels.[\cite{Scanlan1991a}]

There are very few reports concerning rheological measurements of the scaling exponents and properties of critical or near critical colloidal gels as a function of time and moreover, there is a notable absence of reports regarding the critical dynamic behavior captured by $\kappa$. In fact it is not clear that particulate suspensions necessarily traverse a critical state en route to the formation of a colloidal gel.[\cite{KrishnaReddy2012}] The self-similarity of the relaxation modulus originates in the fractal structure of the critical gel. As described above, gelation in colloidal systems often occurs via non-equilibrium routes which may not lead to ideal DLCA or RLCA self-similar structures. Further, the relevance of the fractal concept diminishes at elevated volume fractions where caging or jamming effects become more pronounced, leading eventually to attractive glasses. Finally, power-law relaxation is not unique to critical gels - soft glasses may also display such behavior, as captured phenomenologically by SGR [\cite{sollich1997rheology,sollich1998rheological}], as well as entangled monodisperse polymer melts and other systems.[\cite{Winter1994}] What is unique to the critical gel and the liquid-solid transition however is the inversion of the frequency dependence of $\tan(\delta)$ on crossing $t_g$. In the sol, $\tan(\delta)$ decreases with frequency as expected for a viscoelastic liquid, whereas in the gel $\tan(\delta)$ increases with $\omega$. It is clear that there are several open questions regarding the viscoelasticity of particulate systems during dynamical arrest and that studies to explore these questions are well motivated.

In principle these studies can be advanced readily for particulate gels by following the evolution of frequency dependent dynamic moduli through the gel point as these quantities evolve smoothly through $t_g$, whereas $\eta_0$ and $G_e$ diverge and are difficult to measure in fast evolving systems. In their seminal work, Winter and Chambon made careful observations of viscoelastic properties of gelling polymers by quenching chemical cross-linking reactions at various timepoints of interest.[\cite{Chambon1985,winter1986analysis}] The reversibility of physical gels means that data can be collected from a single sample that is repeatedly cycled through the sol and gel states at different frequencies to assemble a complete map of the dynamic properties of the system, allowing reconstruction of the frequency dependence at any fixed time of interest. Here, we advance such a method to systematically reconstruct the frequency dependence of the dynamic shear moduli as a function of time and temperature for a weakly attractive thermoreversible colloidal gel. The data allow unambiguous identification of the gelation time and reveal its exponential dependence on temperature. We obtain quantitative measurement of the strength ($S_c$) and stress relaxation exponent ($n_c$) for the critical gel, as well as the dynamic exponent $\kappa$. The results here contrast sharply with similar data for a colloidal glass which shows no critical state and a strong dependence of the cross-over time on frequency that cannot be rationalized in the same manner. We conclude that while ergodicity breaking in colloidal gels and glasses bear much in common rheologically from a macroscopic perspective, the dynamical arrest proceeds via very different pathways as revealed by time-resolved characterizations of their viscoelasticity.

\section{\label{experimental}Experimental}
\subsection{\label{materials}Materials}

We work with colloidal silica particles coated with octadecyl chains and suspended in decalin. The particles form a thermoreversible gel below a critical temperature due to the loss of solvency of decalin for the octadecyl chains that provide steric stabilization against aggregation. Particles of diameter $D=90 \pm 6$ nm were prepared using base-catalyzed hydrolysis and condensation of tetraethylorthosilicate (TEOS) as previously described [\cite{ramakrishnan2006microstructure}]. Suspensions were prepared by mechanically mixing a known mass of silica powder in a 50:50 mixture of cis and trans decalin, followed by dilution with additional solvent to obtain the samples with a volume fraction of 17.7\% based on a gravimetrically determined dry mass density of 1.9 g/cm$^3$ for the silica particles. The particles interact with a potential at contact $U/kT=A(T_{\theta}/T-1)$[\cite{jansen1986attractions,rueb1997viscoelastic,ramakrishnan2006microstructure}] where $T_{\theta}$ is the theta temperature for octadecyl chains in decalin and $A$ is a factor proportional to the overlapping volume of the chains in the system. For $\varphi=0.253$ the interaction parameters have been determined from naive mode coupling theory and the temperature dependence of the gel modulus to be $T_{\theta}\approx 327$ K, $A\approx 17$ with a range of attraction given by $\kappa D=11$ where $\kappa^{-1}$ represents a characteristic attraction length in the Yukawa potential. As noted in \cite {ramakrishnan2006microstructure}, although there is an implicit assumption of volume fraction independence in the Yukawa potential, $A$ and $T_{\theta}$ do exhibit volume fraction dependence, although the range of the interaction is largely independent of $\varphi$. Soft colloidal glasses are formed by aqueous suspensions of Laponite charged clay particles at low ionic strength using established procedure.[\cite{Negi2010JoR}] Specifically, 3.5 wt.\% Laponite XLG in water at pH 9.5. Suspensions were vortex mixed for 2 minutes followed by 20 minutes of sonication after which the samples were allowed to stand quiescently for 4 days to allow full hydration of the clay nanoparticles.

\subsection{\label{methods}Methods}

Rheological measurements were conducted in strain controlled mode using using an MCR301 instrument (Anton-Paar) in the cone-plate geometry (1$^{\circ}$, 50 mm diameter steel cone). The low volatility of decalin and the low temperatures used in the experiment resulted in negligible loss of solvent during the experiment so no further attempt to suppress evaporation was necessary. Experiments were conducted by first pre-shearing the samples at a temperature well above the gel-point, 15 $^{\circ}$C at $\dot\gamma=1000 s^{-1}$ for 100 seconds. The temperature was then slowly ramped at 1 $^{\circ}$C/min to the temperature of interest, and the sample was then sheared for 100 seconds at $\dot\gamma=2000 s^{-1}$ followed by cessation of shear using a linear rate ramp of 1 s duration. Measurements of flow curves and time-dependent dynamic moduli were conducted thereafter. This protocol was effective in removing all shear history in the samples and provided a high degree of consistency across multiple runs. For time sweeps, a shear amplitude of $\gamma$=0.5\% was used over a range of frequencies from 40 to 1 rad/s, with data recorded at intervals $\Delta t_m$ of 5 seconds. From a practical perspective measurements at frequencies much lower than 1 rad/s were not pursued as they are the rate of sample evolution defines a lower bound for measurement such that the product of the normalized growth rate of the modulus and the measurement time interval is much smaller than 1, i.e. $(1/G')(\partial G'/\partial t)\Delta t_m\ll 1$. The 5 second interval results in the use of sub-cycle measurement for the lowest frequency of 1 rad/s. Although this resulted in some noisy signals in some cases, the data are robust, and especially so for higher frequencies where the measured data is averaged over several cycles of shear within the 5 second window. Experiments were carried out at 8, 8.5, 9.5, 10, 10.5  and 11 $^{\circ}$C. Temperature control was within $\pm0.1^{\circ}$C or better in all cases. Data are shown for selected temperatures as appropriate. Gelation for 8 and 11 $^{\circ}$C occurred outside the experimental window (before and after, respectively). Laponite suspensions were subjected to a rejuvenating flow of $\dot\gamma$=3000$s^{-1}$ for 100 s then abruptly brought to rest using a linear rate ramp of 1 s duration. Time dependent sweeps were then conducted at a strain amplitude of $\gamma=1\%$ across a range of frequencies $\omega$ from 70 to 1 rad/s.

\section{\label{results_discussion}Results}

\begin{figure}
\includegraphics[width=150mm, scale=1]{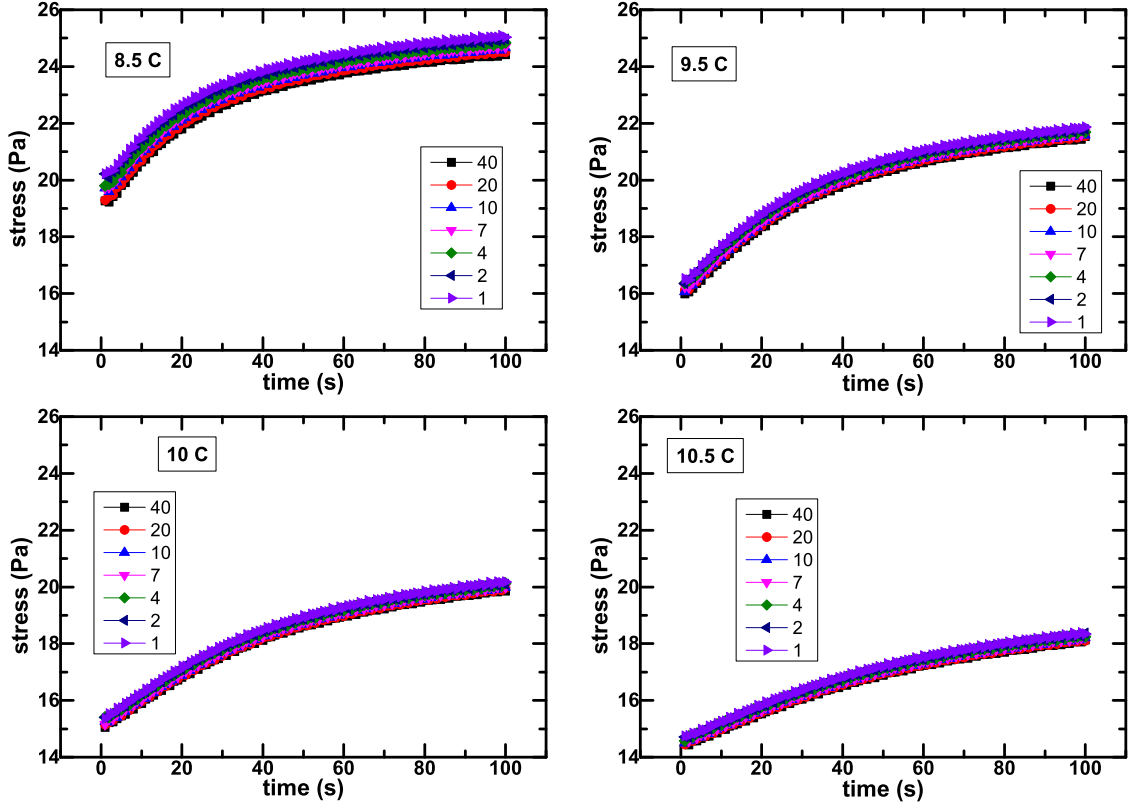}
\caption{Stress evolution during preshear at $\dot\gamma=2000$ s$^{-1}$ before the start of measurements.}
\label{preshear_curves}
\end{figure}

The fidelity of the pre-shear step used in re-initializing gel samples after each run highlights the utility of the protocol used for this investigation. Data are shown in Figure \ref{preshear_curves} for the pre-shear steps at $\dot\gamma=2000s^{-1}$associated with subsequent time sweeps across the different frequencies indicated, for 4 selected temperatures, 8.5, 9.5, 10 and 10.5 $^{\circ}$C. While the curves do not asymptote fully to a steady state value, the measured viscosities after 100 s of shear exhibit minimal differences. This serves to establish a reproducible initial condition for subsequent time sweeps. The sensitivity of the system to temperature is evident with a \textit{ca}. 25\% decrease in the viscosity on increasing the temperature to 10.5 from 8.5 $^{\circ}$C.

\begin{figure}
\includegraphics[width=150mm, scale=1]{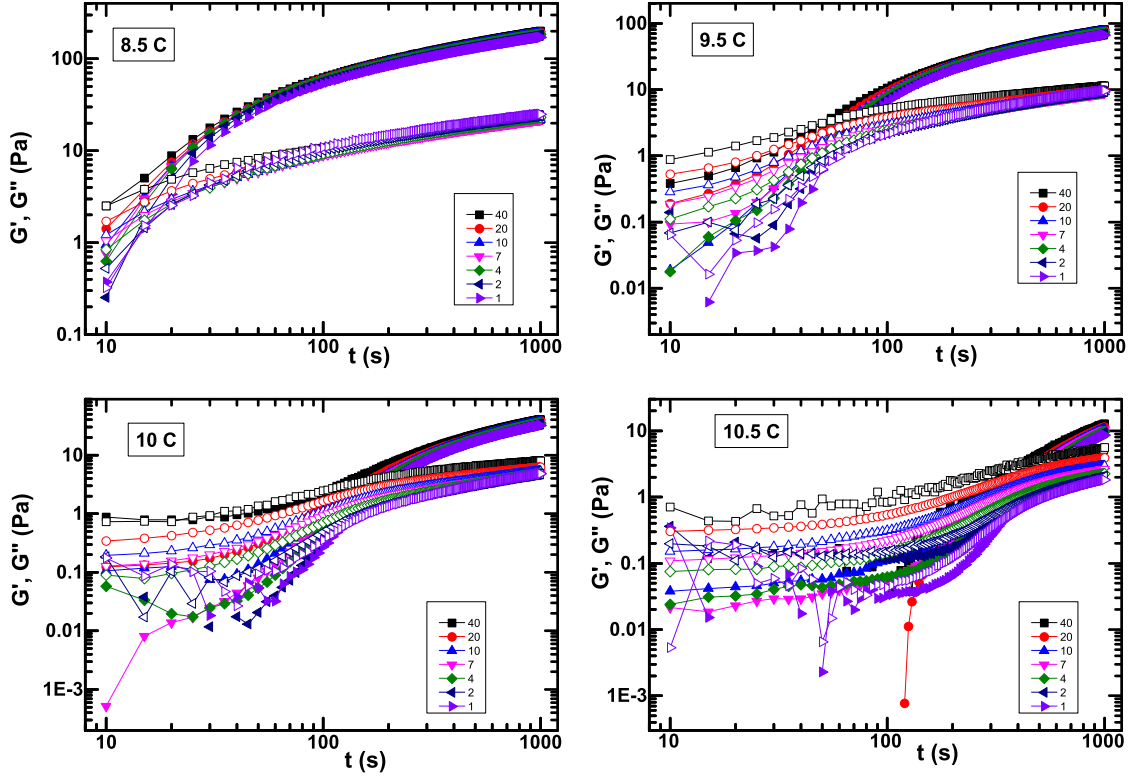}
\caption{Time evolution of dynamic moduli at different frequencies, for different temperatures as indicated.}
\label{modulus_time}
\end{figure}

\begin{figure}
\includegraphics[width=150mm, scale=1]{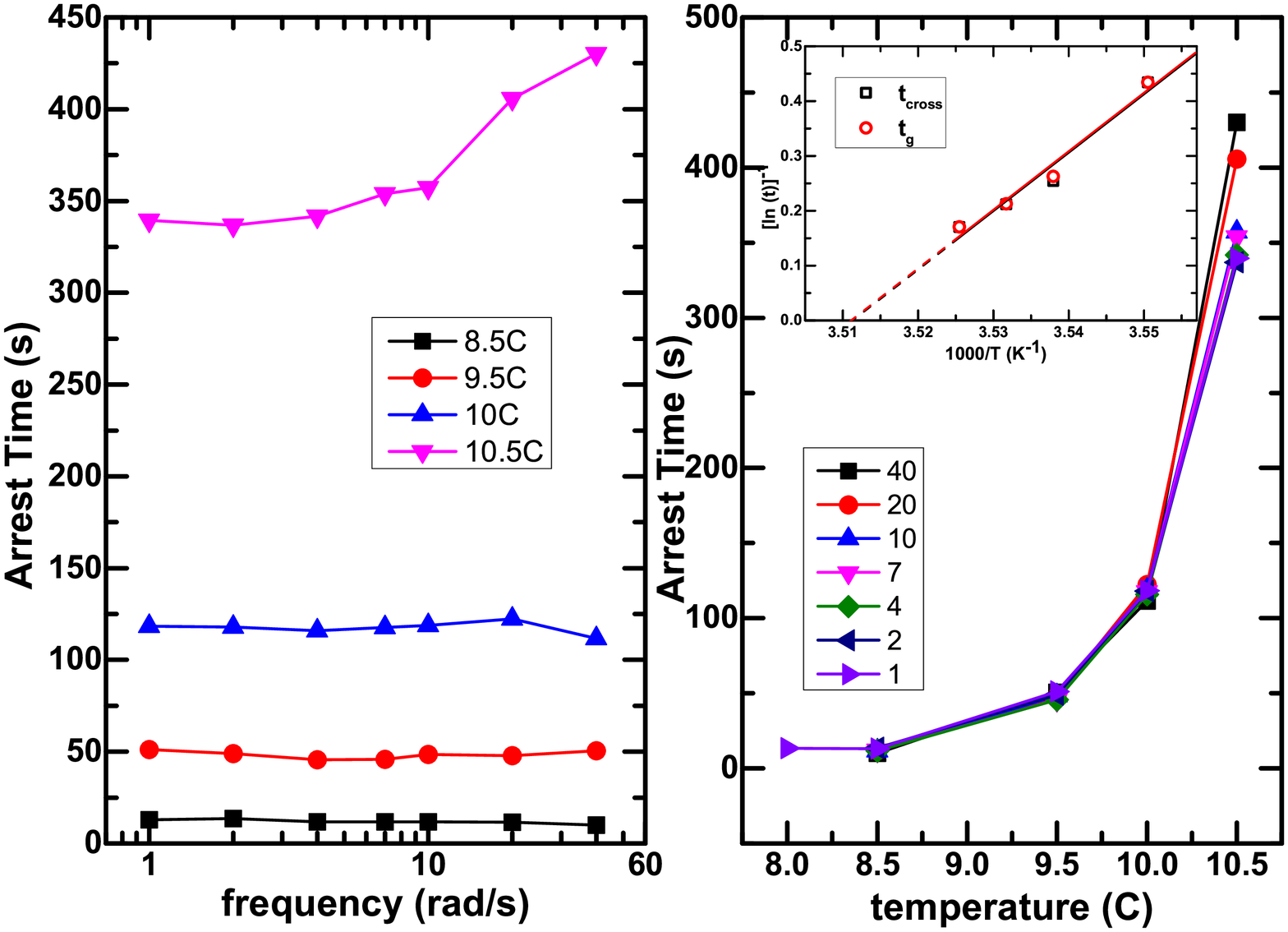}
\caption{Left: Frequency dependence of the arrest time for different temperatures as indicated. Right: Temperature dependence of the arrest time. Extrapolation to zero of $1/\log(t_g)$ and $1/\log(t_{cross})$ yield virtually indistinguishable estimates for the gelation temperature of 11.74 and 11.72 $^{\circ}$C, inset.}
\label{arrest_time_freq_temp_dependence}
\end{figure}

Figure \ref{modulus_time} provides time traces of elastic and viscous moduli across different frequencies, at selected temperatures as indicated. In all cases, the rate of increase of the elastic modulus exceeds that of the viscous modulus, with both quantities showing a power law evolution at the longest times considered. For the elastic modulus, $G'\sim t^{\alpha}$ where $\alpha$ changes from 0.4 to 0.6, 0.87 and 1.73 as temperature is increased from 8.5 to 10.5 $^{\circ}$C. At sufficiently short times, $G''>G'$ across all frequencies but there is a crossover between the two after a characteristic arrest time or crossover time $t_{cross}$. Notably, $t_{cross}$ is only weakly dependent on frequency, with crossover occurring over a very small range of times in all cases. If we associate $t_{cross}$ as the time at which the system has a relaxation time given by $\tau=1/\omega$ we arrive at the peculiar result that the system possesses all relaxation times at $t=t_{cross}$ and that the longest relaxation time is infinite, Figure \ref{arrest_time_freq_temp_dependence}. This succinctly reflects the critical nature of the gel transition. The size distribution of clusters is infinitely broad at gelation, resulting in a material that displays all possible relaxation time scales.[\cite{Rueb_Zukoski1998}] The slight upturn for higher frequencies for the 10.5 $^{\circ}$C sample may be due to inadverdent inertial effects which are more prominent at higher frequencies in such weak gels.

\begin{figure}
\includegraphics[width=150mm, scale=1]{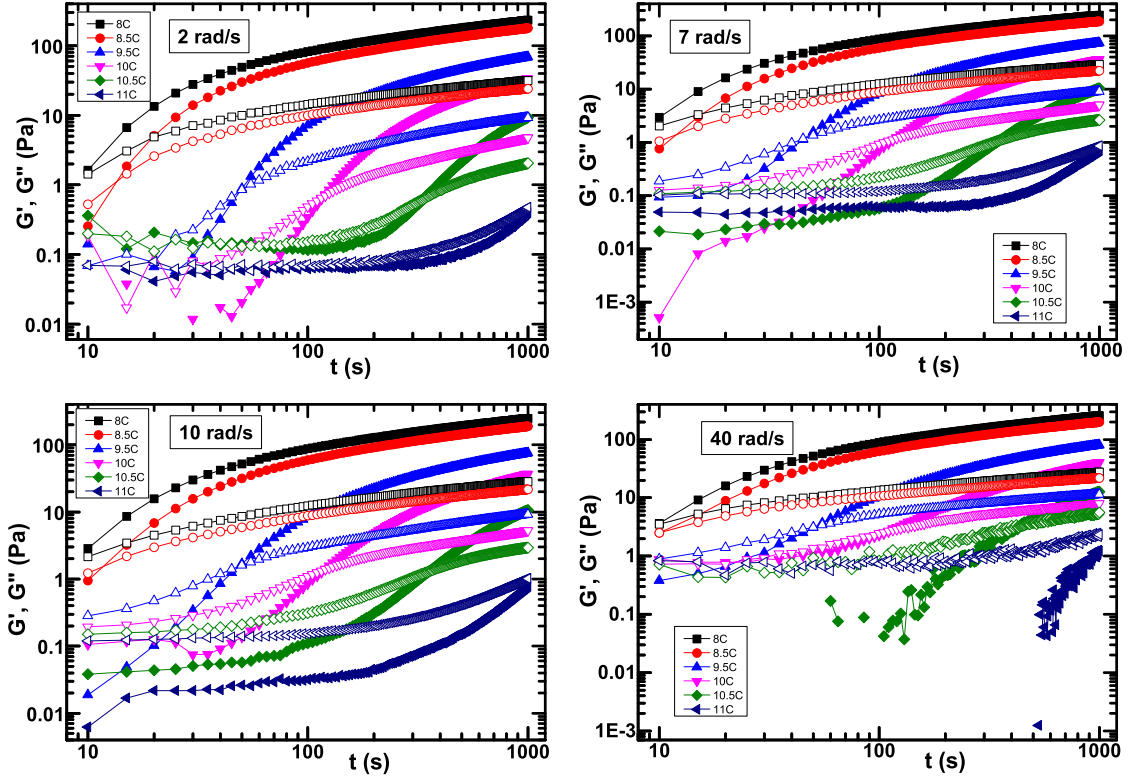}
\caption{Evolution of dynamic moduli across different temperatures, for constant frequencies as indicated. The crossover between G' and G'' marks the arrest time or $t_{cross}$ which depends very sensitively on temperature.}
\label{arrest_time_temp_dependence}
\end{figure}

The system exhibits a pronounced dependence of arrest time on temperature. As shown in Figures \ref {arrest_time_temp_dependence} and \ref{arrest_time_freq_temp_dependence}, at a fixed frequency, $t_{cross}$ varies strongly with temperature, ranging from $\approx$ 10 s at 8.5 $^{\circ}$C to over 900 s at 11 $^{\circ}$C. This indicates that the system is in the RLCA regime since the probability of forming a bond between particles and aggregates and eventually the gel is a strong function of the attraction strength. Conversely, the modulus of the gel at cross-over displays a very weak dependence on temperature. At $\omega$=10 rad/s for example, $G^*$ changes from $\approx$ 1.5 to 1.2 Pa from 8.5 to 10.5 $^{\circ}$C. This dependence is slightly greater at lower frequencies, but in all cases, the moduli at crossover are roughly identical, within a factor of 1.5-2. The strong similarity across the shear moduli despite the large differences in the aggregation kinetics is striking. The gel temperature may be estimated by extrapolating the temperature dependence of $[\ln(t_{cross})]^{-1}$ and $[\ln(t_g)]^{-1}$ to zero. The extraction of $t_g$ is described later. As expected, given the lack of dependence of $t_{cross}$ on frequency, extrapolation based on $t_{cross}$ and $t_g$ yield nearly indistinguishable estimates of 11.72 and 11.74 $^{\circ}$C, respectively for $T_g$. Based on the model for the short range attraction between the octadecyl brush layers, the interaction strength changes by just 0.21 $kT$ from 2.78 at 8 $^{\circ}$C to 2.57 at 11 $^{\circ}$C, and at the gel temperature, $U/kT$ is 2.52. Using $T_g=11.75 ^{\circ}$C, we estimate the relative increase in the gel modulus on decreasing temperature from to 8 $^{\circ}$C from 11 $^{\circ}$C using the naive mode coupling result which predicts $G'/kT\sim\exp[AT_{\theta}(1/T-1/T_g)]$ [\cite{ramakrishnan2006microstructure}], leading to $G'_{281 K}/G'_{284 K}=1.23$. Alternatively, the simple exponential dependence advocated by [\cite{rueb1997viscoelastic}] yields $G'_{281 K}/G'_{284 K}=1.17$. The proximity of these samples to $T_g$ results in very similar shear moduli, whereas the small difference in the interaction potential has a much sharper effect on the gelation time, likely due to a non-trivial influence on the sticking probability, of which $t_g$ is a sensitive function.[\cite{Schmitt2000,Runkana2005}]

\begin{figure}
\includegraphics[width=150mm, scale=1]{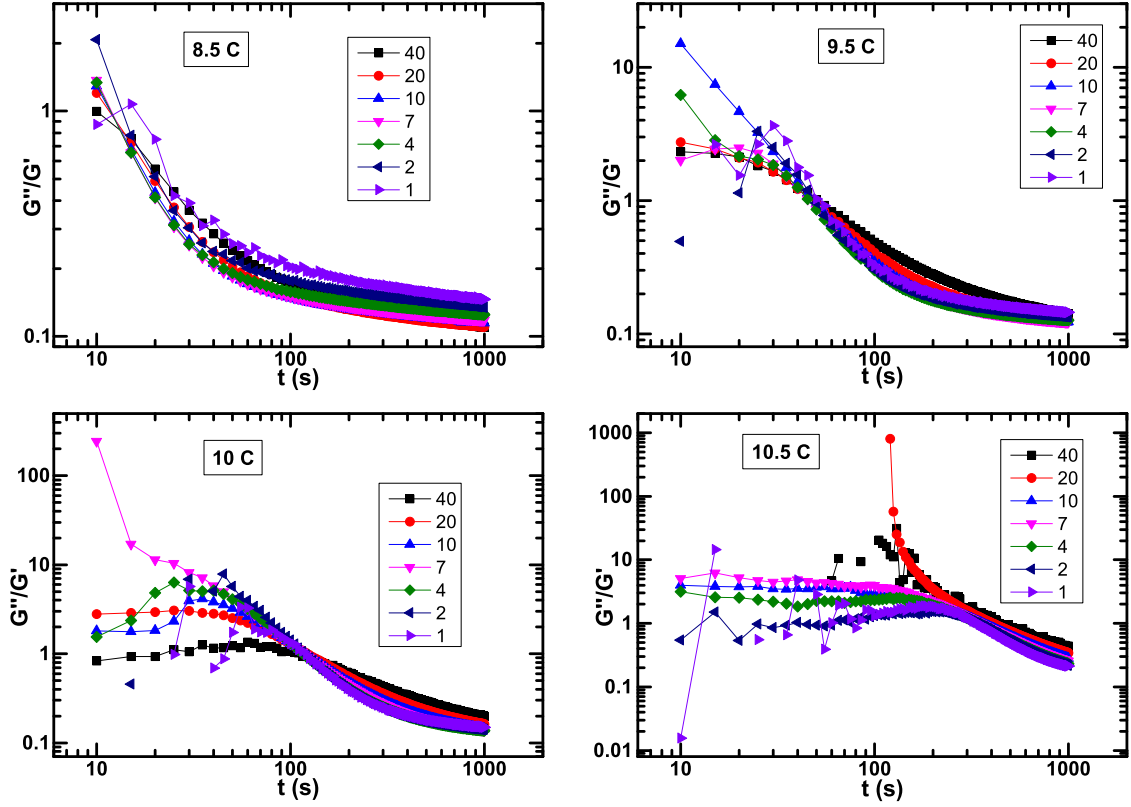}
\caption{Evolution of $\tan[\delta(t)]=G''/G'$ across different frequencies, for different temperatures as indicated. The gel time $t_g$ is marked by the intersection of the $\tan(\delta(t))$ curves at $\approx$ 45, 110 and 350 seconds for 9.5, 10 and 10.5 $^{\circ}$C, respectively. Equivalently, at this time, $\tan(\delta)$ is independent of frequency.}
\label{tan_delta_time}
\end{figure}

\begin{figure}
\includegraphics[width=150mm, scale=1]{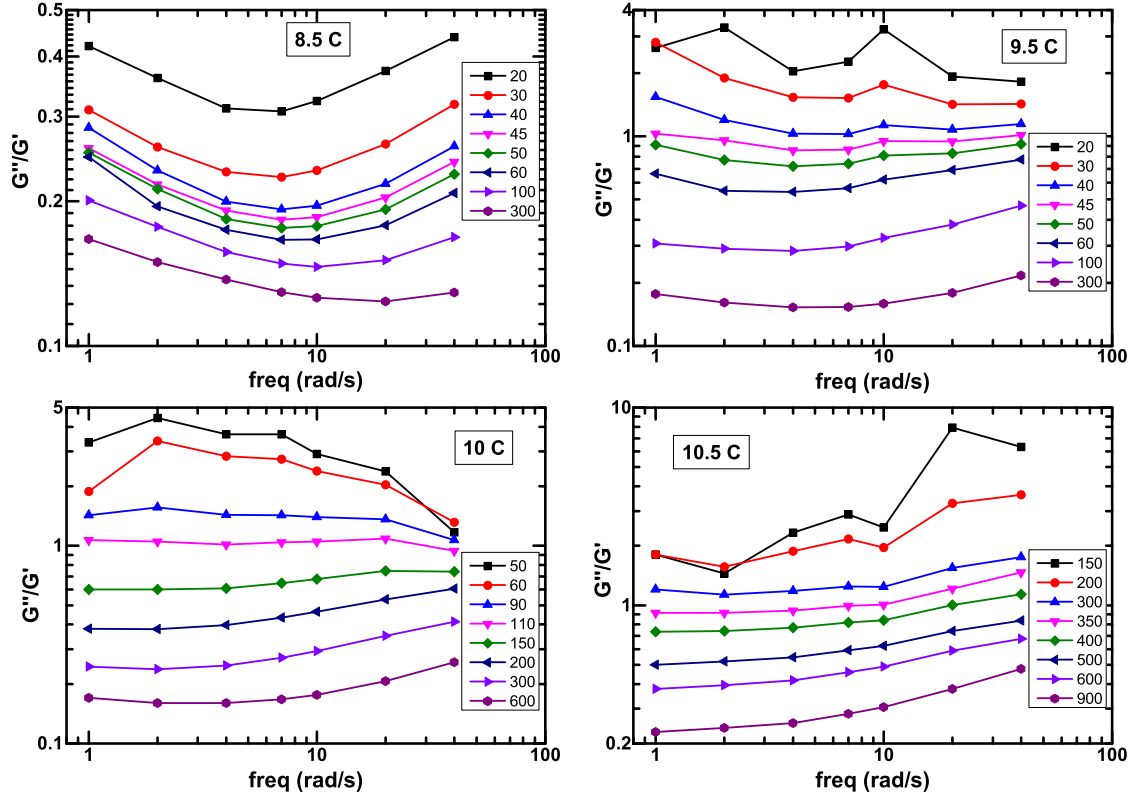}
\caption{Frequency dependence of the loss angle, $\tan[\delta(\omega)]=G''/G'$ across different timepoints, for temperature (from left to right) of 8.5, 9.5, 10.0 and 10.5 $^{\circ}$C. The gel time, $t_g$ is the time at which $\tan[\delta(\omega)]\sim\omega^0$. This occurs approximately at 45, 110 and 350 seconds for 9.5, 10 and 10.5 $^{\circ}$C, respectively.}
\label{tan_delta_freq}
\end{figure}

We plot the time and frequency dependence of $\tan(\delta)$ and determine the gel time as that time at which there is intersection among the traces at different frequencies and the time at which $\tan(\delta)\sim\omega^0$, respectively. Data are shown in Figures \ref{tan_delta_time} and \ref{tan_delta_freq}. The time dependent traces (Figure \ref{tan_delta_time}) show that the data for different frequencies intersect at fairly well defined timepoints for each sample temperature. The representation in $\tan[\delta(\omega)]$ is a bit less clear, but still enables estimation of a timepoint at which the frequency independence of the loss angle can be established. That is, the timepoint at which the frequency dependence of the loss tangent inverts from a decreasing function of $\omega$ for the liquid-like viscoelastic sol to an increasing function of the solid-like viscoelastic gel. The data for the 10 $^{\circ}$ sample is particularly striking in this regard. The scatter in the data at large loss angle is simply due to the low torque from measurements on the low viscosity sol. The lack of strict coherence with the form schematically shown in Figure \ref{tan_delta_gelation_theory} likely originates from inhomogeneities in the sample [\cite{Scanlan1991a}] which smear out the response of the critical gel. This may be caused by unavoidable local temperature gradients or fluctuations in the atmosphere-exposed cone-plate geometry. We estimate $n_c$, the critical relaxation exponent, from the numerical value of the loss tangent at the identified gelation time. The resulting values varied from 0.47 to 0.52, which suggests that here $n_c\approx 0.5$ is a reasonable representation of the dynamics of the system for all the temperatures considered. The similarity in the time dependence with the expected form schematically shown in Figure \ref{tan_delta_gelation_theory} is noteworthy. We determine the gel times for 9.5, 10 and 10.5 $^{\circ}$C as \textit{ca}. 45, 110 and 350 seconds, respectively. For 8.5 $^{\circ}$ C it is apparent that gelation occurs quickly after the start of the measurement, and we estimate the gelation time to be \textit{ca}. 10 s, but this is subject to a large uncertainty given the small number of data points in this short time regime.

\begin{figure}
\includegraphics[width=150mm, scale=1]{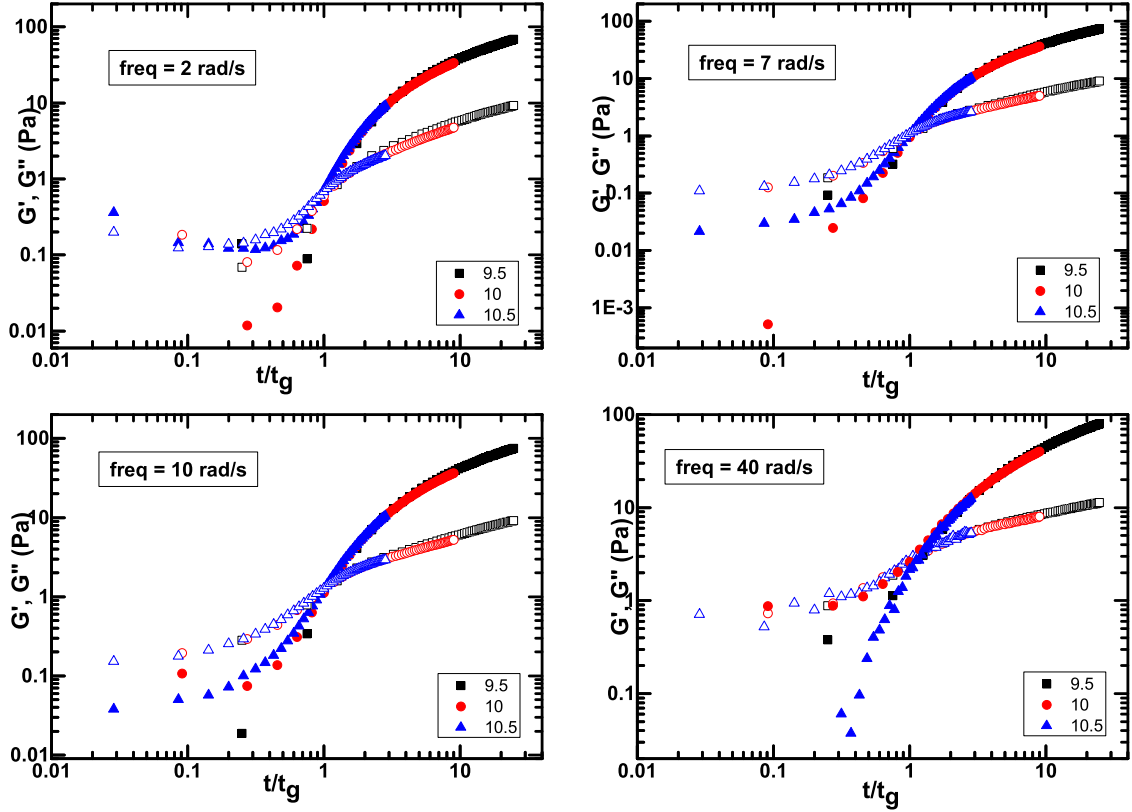}
\caption{Evolution of dynamic moduli across different temperatures, at constant frequencies (from left to right) of 2, 7, 10 and 40 rad/s. The data all fall on a single curve when experimental time is rescaled by the gel time, $t_g$ indicating that the systems evolve along a single trajectory but with different kinetics that are accurately captured by a single scaling parameter, $t_g$.}
\label{arrest_time_temp_dependence_rescaled}
\end{figure}

We use the extracted gel times to further examine the evolution of the viscoelasticity during gelation. As seen in Figure \ref{arrest_time_temp_dependence_rescaled}, the data for different temperatures at a given frequency can be rescaled onto a single master curve using an effective time $t/t_g$. The similarities in the trajectories of the samples are striking, and the data suggest that the intrinsic modulus of these gels, $G'(t=\infty)$, is rather insensitive to temperature, for the small range of temperatures considered here. This can be rationalized by considering the very small change in the inter-particle potential from 2.68 to 2.61 $kT$ from 9.5 to 10.5 $^{\circ}$C, despite the strong change in aggregation kinetics. The data can be fit using a first-order model to describe the increase of the modulus, $G'=G'_{\infty}\left(1-\exp\left[-\alpha(t/t_g-1)\right]\right)$, where $\alpha$ represents the rate of increase of connectedness of the gel.[\cite{rueb1997viscoelastic}] This procedure yields an estimate for $G'_{\infty}$ that displays a weak power-law dependence, $\sim\omega^{0.05-0.1}$ with $G'_{\infty}\approx$ 100 Pa at $\omega$=40 rad/s., with $\alpha$ roughly constant at 0.07. The lack of a rigorous microscopic interpretation of $\alpha$ prevents us from further leveraging this analysis. Moreover, the sensitivity of the model to the diverging modulus at short times limits the utility of this approach for the long experimental timescales here, $t/t_g\approx 30$. Nevertheless it is clear that $t_g$ as a single parameter is sufficient to describe the trajectories of these samples as they gel and continue to evolve over time as also observed in prior work on silica gels. [\cite{rueb1997viscoelastic,Cao2010}] We also note that the temperature independence of the rate of increase of gel connectivity and exponential temperature dependence of $t_g$ in close proximity to the gel point has also been recovered by Leheny et al.[\cite{Guo_Leheny2011}]


\begin{figure}
\includegraphics[width=150mm, scale=1]{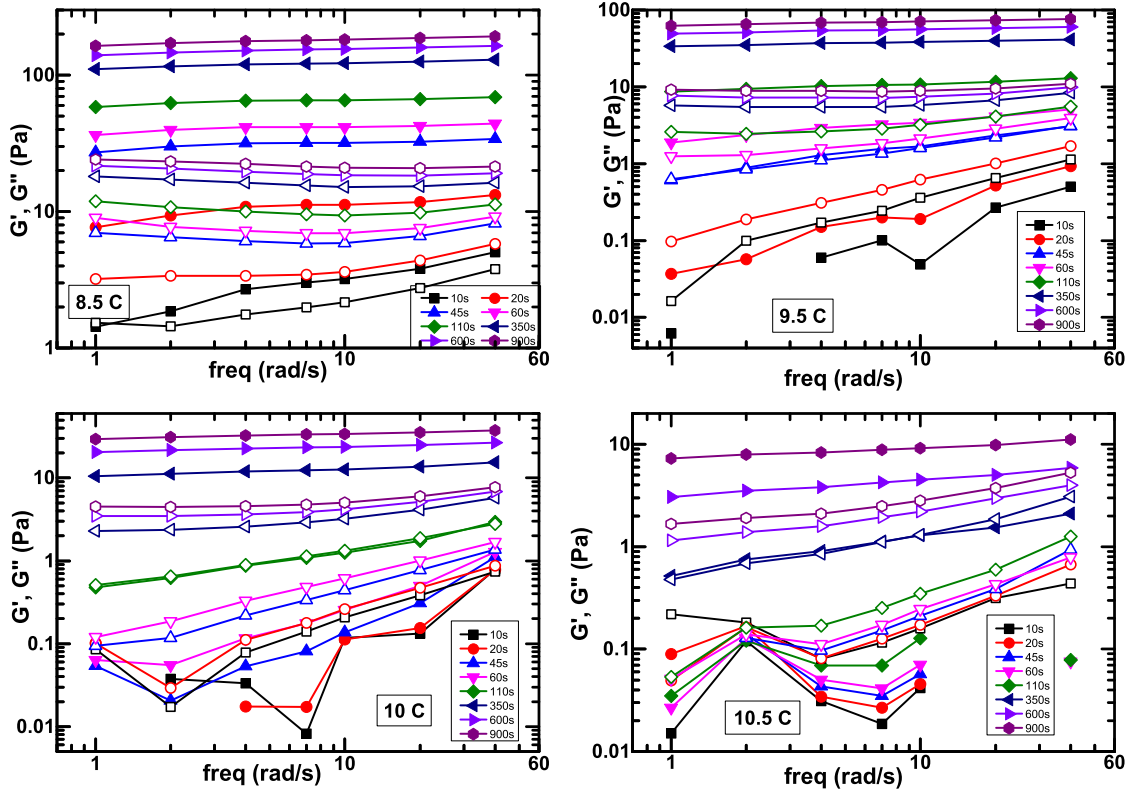}
\caption{Frequency dependence of the storage and loss moduli across different times during gelation, for different temperatures, as indicated. The near critical gel is apparent at $t\approx$ 45, 110 and 330 seconds for 9.5, 10 and 10.5 $^{\circ}$C, respectively.}
\label{modulus_freq_dependence_temperatures}
\end{figure}

\begin{figure}
\includegraphics[width=150mm, scale=1]{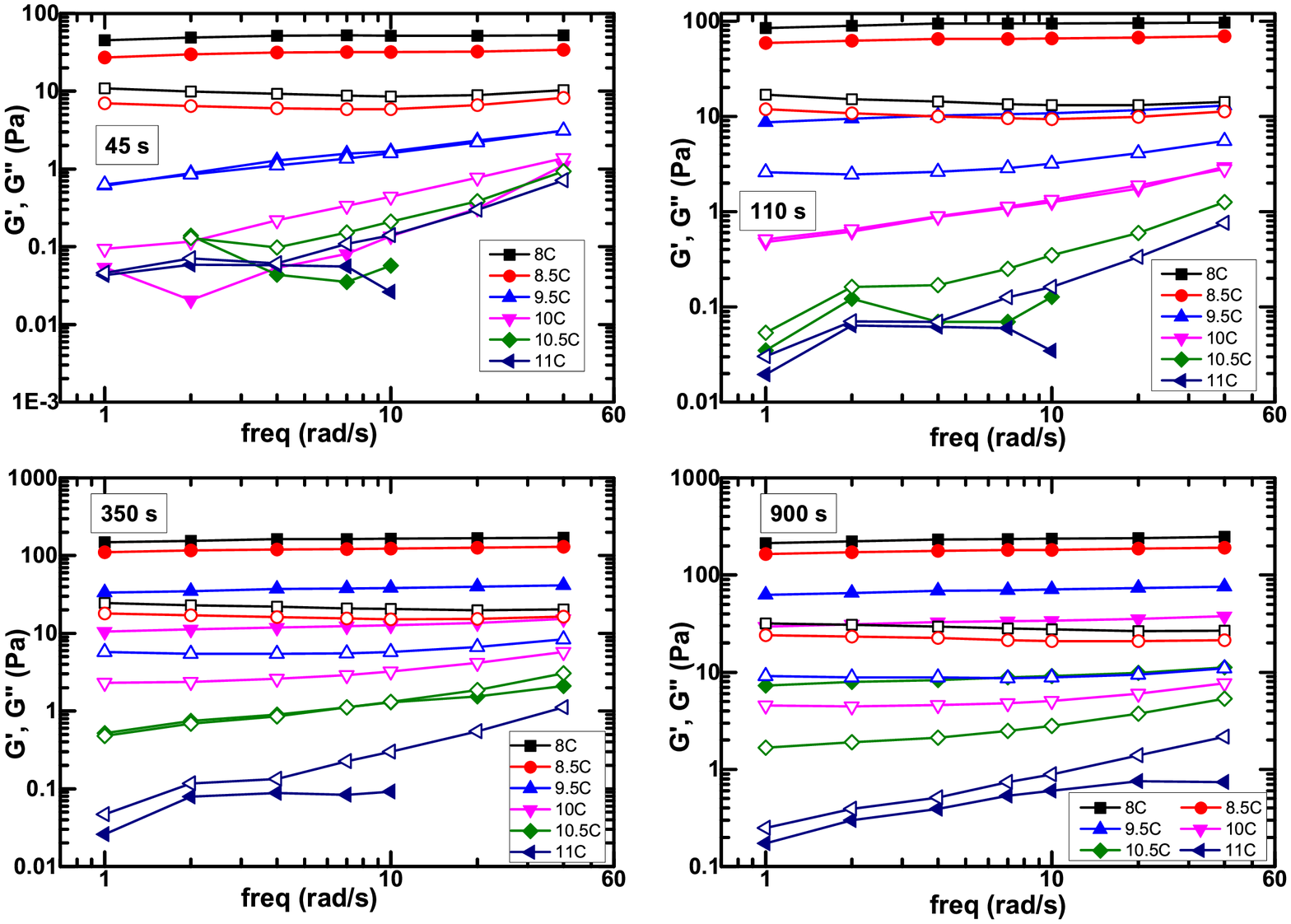}
\caption{Frequency dependence of the storage and loss moduli across different temperatures, for different times, as indicated. The near-critical gel state is apparent at \textit{ca.} 45, 110, 350 seconds for 9.5, 10 and 10.5 $^{\circ}$C, respectively. The sample at 11 $^{\circ}$C is just shy of the critical state at 900 seconds.}
\label{modulus_freq_dependence_times}
\end{figure}

The frequency dependent moduli at different fixed times provide additional perspective on the evolution of the viscoelasticity during gelation, Figure \ref{modulus_freq_dependence_temperatures}. The long time behavior is immediately recognizable as that of a solid-like material, with $G'>G''$ and zero or weak dependence of $G'$ on frequency. At short times, $G''>G'$ and both quantities display a marked frequency dependence. There exists a critical time for each temperature or interaction strength where $G'\approx G''\sim\omega^n$. This occurs at roughly 45, 110 and 350 seconds for 9.5, 10 and 10.5 $^{\circ}$C respectively, with $n\approx0.5$ in  all three cases. We associate this timepoint with the critical gel, and so we consider $n_c$=0.5, consistent with the value deduced from the analysis of the loss tangent. Notably, there is no time point for which there is a clear indication that the system is ``liquid-like'', $G''>G'$ at lower frequencies and ``solid-like'', $G'>G''$ at higher frequencies. That is, there is no timepoint at which one can identify a crossover between $G'$ and $G''$ in frequency. The elastic and viscous moduli move towards each other over time, achieve congruence at a critical time, and then move apart without intersection in $\omega$-space, that is, with $G'>G''$ for all frequencies and times beyond the critical time. This indicates that although the sample is not a chemical gel, the inter-particle interactions are strong enough to place the relaxation time of the system beyond the experimental frequency range. The congruence of the elastic and viscous moduli and the power-law scaling with frequency indicate self-similar relaxation, which is expected for the critical gel. The frequency dependence at different temperatures, for selected times provides a complementary perspective, Figure \ref{modulus_freq_dependence_times}. In the context of Equation \ref{eq:dynamic_moduli_critical_gel}, the critical relaxation exponent of $n_c\approx0.5$ explains why the cross-over time provides a fairly accurate measure of gelation time, and why it is effectively independent of frequency as $G'$ and $G''$ are scaled by the same factor, $\cos(n_c\pi/2)=\sin(n_c\pi/2)=1/\sqrt{2}$ for $n_c=0.5$. From the frequency dependence of the critical gel we yield a quantitative measurement of $S_c\approx 0.33$, using Equation \ref{eq:dynamic_moduli_critical_gel}. A similar finding of $n_c=0.5$ has been reported recently by Eberle et al. for octadecyl silica colloids in tetradecane.[\cite{Eberle2011}]

\begin{figure}
\includegraphics[width=150mm, scale=1]{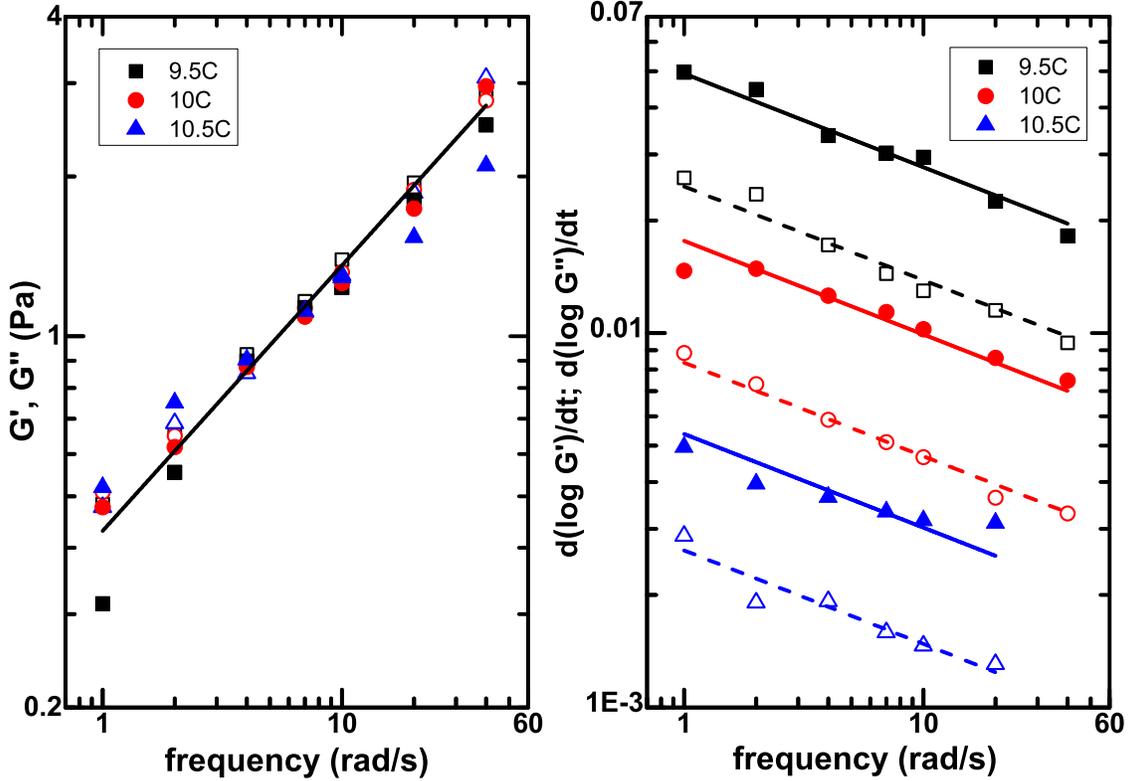}
\caption{Left: Storage and loss modulus in the near critical gel for 9.5, 10 and 10.5 $^{\circ}$C. $G'\sim G'' \sim\omega^{0.5}$. A line of slope 0.5 is drawn as a guide to the eye. Right: Rate of change of the modulus as a function of frequency at the identified gel points of $t_g$ =45, 110 and 350 s for 9.5, 10 and 10.5 $^{\circ}$C. Lines of slope $\kappa=0.25$ are drawn as guides for the eye. Data conform to the relationship shown in Equation \ref{eq:modulus_rate} with a proportionality of $C\approx$ 2.}
\label{critical_gels_aging_rate}
\end{figure}

Close inspection of the timepoints in Figures \ref{modulus_freq_dependence_temperatures} and \ref{modulus_freq_dependence_times} where $G'\approx G''\sim\omega^{0.5}$ suggests that the critical gels not only have the same frequency dependent viscoelasticity, but that the magnitudes of the moduli are similar. Data for 9.5, 10 and 10.5 $^{\circ}$C at their gel times are clearly resolvable, and are plotted separately in Figure \ref{critical_gels_aging_rate}, where their near-overlap is readily apparent. As clear picture emerges of a system which transitions from a liquid-like to a solid-like behavior via a common critical state, but which arrives at this critical state at different times due to the strong temperature dependence of the aggregation kinetics. The critical gel is described completely by its relaxation modulus, $G(t)=0.33t^{-0.5}$. Finally, we consider the frequency dependence of the rate of evolution of the critical gel. As shown in Figure \ref{critical_gels_aging_rate}, for 9.5, 10 and 10.5 $^{\circ}$C, the system exhibits a weak power law dependence of the modulus growth rate on frequency, with a dynamic critical exponent (Equation \ref{eq:modulus_rate}) $\kappa\approx$ 0.25. This value is similar to that found in reports on polymer gels where Winter et al. observed $\kappa=0.21\pm 0.02$ for a PDMS gel [\cite{Scanlan1991}], Colby et al. observed $\kappa=1/(vz)=0.24$ for a polyester gel, [\cite{colby1993dynamics}] and $\kappa=0.25$ was found for epoxy networks.[\cite{Eloundou1996,Mortimer2001}] Additionally, we observe that the rate of evolution of the elastic modulus is roughly twice that of the viscous modulus, $C\approx 2$, Equation \ref{eq:modulus_rate}, across the entire frequency range, as also witnessed for a PDMS gel.[\cite{Scanlan1991}]

\begin{equation}
\left(\frac{1}{G'}\frac{\partial G'}{\partial t}\right)_{t=t_g}\approxeq C \left(\frac{1}{G''}\frac{\partial G''}{\partial t}\right)_{t=t_g} \sim \omega^{-\kappa}
\label{eq:modulus_rate}
\end{equation}

\begin{figure}
\includegraphics[width=100mm, scale=1]{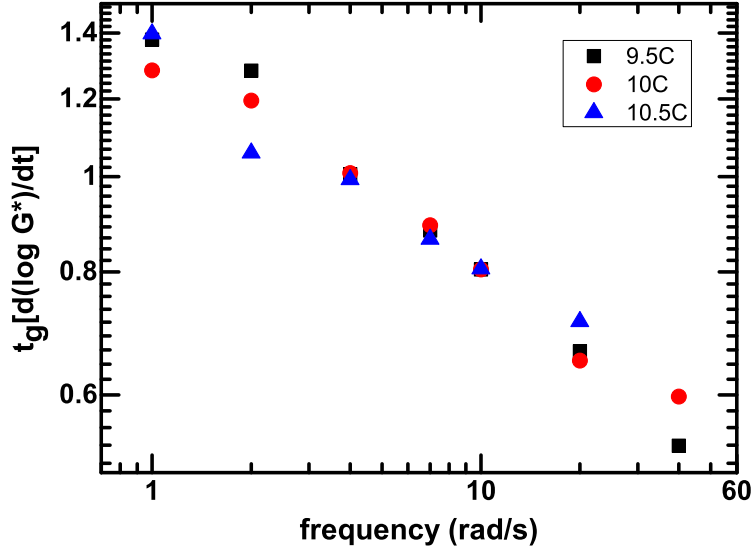}
\caption{Dimensionless product of the gel time and aging rate for 9.5, 10 and 10.5 $^{\circ}$C as indicated.}
\label{critical_aging_universal}
\end{figure}

Winter and Scanlan have suggested that the dynamic critical exponent adopts a universal value for gelation. Although at present we do not have a clear indication that $\kappa$ is a universal exponent, the observation of $\kappa\approx 0.25$ here for a physical particulate gel is notable by comparison with available data from polymer gels. With knowledge of $\kappa$ and $n_c$ provided directly by measurements, we use the result of scaling arguments to obtain $k$ and $z$ describing the divergence of the equilibrium properties leading up to and moving away from the critical gel. Using Equation \ref{eq:exponents} we see that $k=(1-n_c)/\kappa=2$ and $z=n_c/\kappa=2$. The data of Figure \ref{critical_gels_aging_rate} provide a characteristic rate for the evolution of the modulus as a function of frequency for different temperatures, with the system showing faster evolution at lower temperatures where inter-particle attraction is stronger. The significance of the rates themselves is unclear. We may speculate however that these rates should exhibit universal behavior if normalized by an appropriate characteristic time. The only temperature dependent characteristic timescale in the system is the gelation time $t_g$, and so we plot the product of $t_g$ and $[\partial (\log G^*)/\partial t]_{t_g}$, Figure \ref{critical_aging_universal}. The data indeed do demonstrate a robust collapse, for all 3 temperatures. The specific reasons for this are currently unclear, but it seems reasonable to interpret the universal behavior as connected to the dimensionless $\alpha$ in the first-order kinetic model that captures the evolution of the gel as a function of effective time $t/t_g$. To the best of our knowledge, there have been no prior reports of measurements of the dynamic critical exponent for particulate gels and of universal behavior in the rate of evolution of the complex modulus of critical gels when scaled by $t_g$.

\begin{figure}
\includegraphics[width=150mm, scale=1]{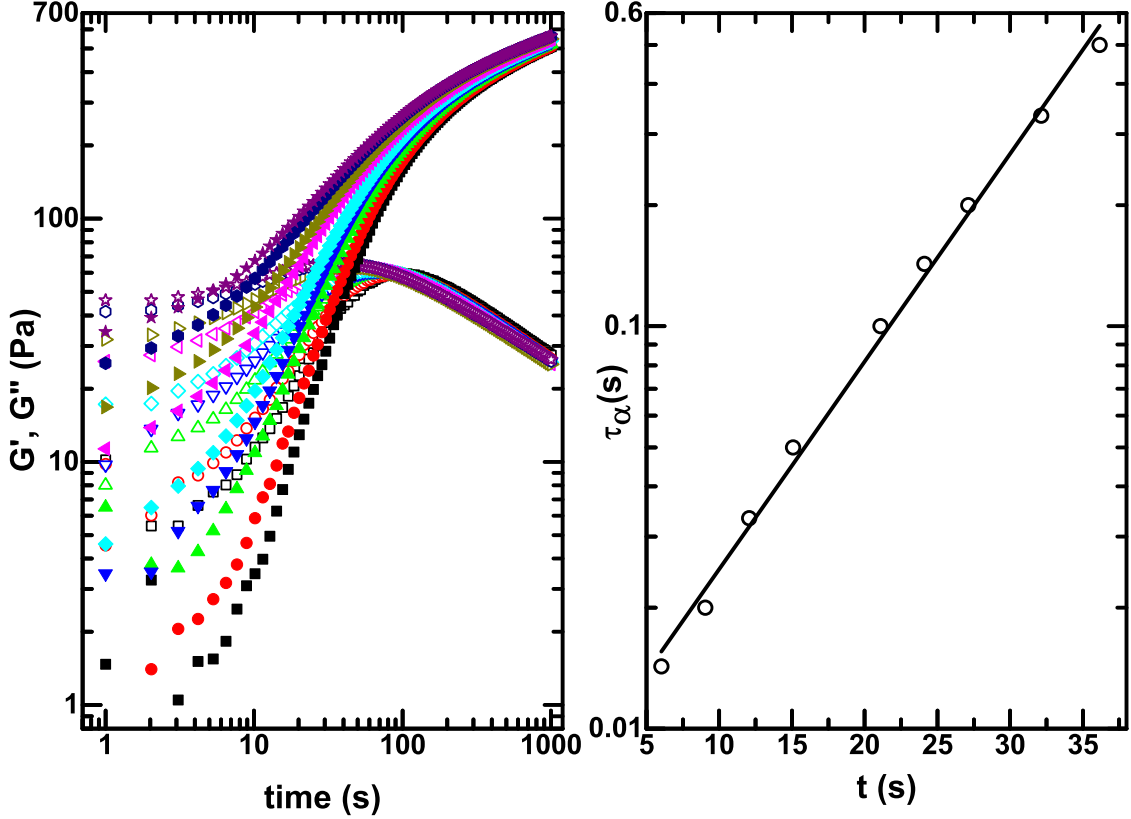}
\caption{Left: Time evolution of $G'$ (closed symbols) and $G''$ (open symbols) for laponite glass at frequencies from top to bottom of  of $\omega$=70, 50, 30, 20, 10, 7, 5, 3 and 2 rad/s. Right: Aging of the relaxation time $\tau_{\alpha}=1/\omega$ at $t=t_{cross}$.}
\label{laponite_modulus_evolution}
\end{figure}

The preceding data suggests quite strongly that gelation in colloidal suspensions reflects much of the character of gelation in crosslinking polymers. As such, gelation occurs with passage through an identifiable critical gel state, the properties of which have been subjected to direct measurement. Colloidal glasses represent an interesting juxtaposition with colloidal gels. The evolution of the shear modulus as a function of time on first glance is strikingly similar similar to that for gels - there is a cross-over between $G'$ and $G''$ , with power-law evolution at long times, and an abrupt upturn of the elastic modulus at short times near the cross-over, particularly for lower frequencies. When we subject this data to the same frequency resolved treatment as done for the colloidal gel however, the differences between the two apparently rheologically similar systems begin to emerge. We consider data here for a colloidal glass formed by repulsive electrostatic interactions between Laponite clay particles in low ionic strength aqueous suspension, a commonly studied system.[\cite{Bonn2002,Joshi2008,Joshi2012}] The crossover time shows a rather more marked frequency dependence such that crossover time shows an inverse logarithmic dependence on frequency. Viewed another way, the system has a relaxation time equal to the inverse of the sampling frequency when the age of the system is equal to the crossover time. From this perspective, we see that the relaxation time of the system evolves exponentially with time, Figure \ref{laponite_modulus_evolution}. This interpretation has been substantiated by studies of stress-mediated aging in Laponite glasses.[\cite{Negi2010,Negi2010EPL}] The logarithmic frequency dependence implies that over a typical rheological frequency window, the crossover time can change by a factor of 2-3. It was not possible to bring the cross-over times into registry by normalizing $G'(t)$ and $G''(t)$ by $\cos(n_c\pi/2)$ and $\sin(n_c\pi/2)$ for $0<n_c<1$. This highlights the fact that the structural features associated with dynamical arrest in the glass are qualitatively different than those in the gel, \textit{viz}. cage formation versus system-spanning particle bonding or cluster jamming. In the glass there appears to be a continual slowing down of the system over time as particles become caged, but mobility through collective motion on longer length scales (low frequencies) is maintained.

\begin{figure}
\includegraphics[width=150mm, scale=1]{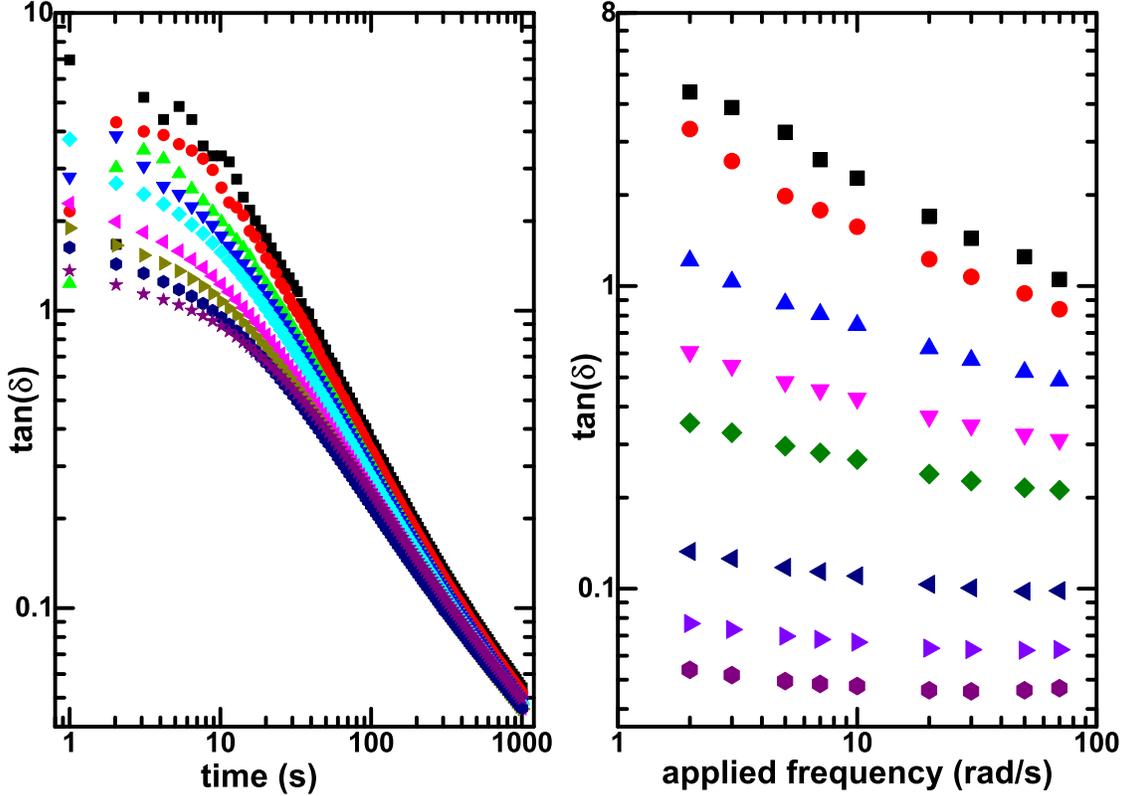}
\caption{Left: Evolution of $\tan(\delta)$ as a function of time for the $\omega$= 70, 50, 30, 20, 10, 7, 5, 3 and 2 rad/s. Right: Frequency dependence of $\tan(\delta)$ at various times, from top to bottom, $t$=4, 10, 30, 60, 100, 300, 600 and 1000 seconds.}
\label{laponite_tan_delta_time_freq}
\end{figure}

Aside from the above heuristic discussion, an examination of the time and frequency dependence of the loss tangent reveals pronounced differences in the dynamical arrest of the glass. As shown in Figure \ref{laponite_tan_delta_time_freq}, there is no intersection in time among the curves for different frequencies, and no delineation of liquid-like from solid-like states in the frequency dependence for different times. That is, there is no inversion in the frequency dependence of the loss tangent, as would be expected in a fluid-to-solid transition in a conventional viscoelastic material. The evolution of the glass occurs without passage through any identifiable critical intermediate that delineates the pre- and post-arrest states. Instead, there is a continuous transition with $\tan(\delta)$ progressively decreasing as the relaxation time of the system increases with time after cessation of flow. The relaxation time defined as the inverse of the sampling frequency in fact provides the means to rescale the frequency dependent properties onto a universal master curve that displays a well-defined terminal regime. By contrast, for a colloidal gel, formally defined, the relaxation time of the system is not finite for $t>t_g$. This is strictly true for chemical gels, but also pertains for physical gels where the bond lifetime does not subject the system to creep over the experimental timeframe. Recent work by Winter addresses the contrast between gelation and vitrification in some detail, with an effort devoted to discriminating between colloidal gels and colloidal glasses based on the form of their relaxation spectra. Both are power-law functions, but for colloidal glasses, $H(\tau)$ has a positive exponent whereas for colloidal glasses the exponent is negative.[\cite{Winter2013}]

\section{Discussion}

Understanding the nature of the fluid-to-solid transition in colloidal suspensions has clear implications for a broad range of complex fluids. Dilute systems with strong attractions which aggregate irreversibly and dense systems with hard sphere interactions represent limiting cases which have been well described. The intervening regimes of both volume fraction and attraction strength present a rich variety of behavior [\cite{Negi2009,Shao2013}], but with somewhat universal features as discussed by [\cite{Prasad2003}]. Of note is the observed power-law divergence of relaxation time in the vicinity of the onset of solid-like properties, $\tau_{\alpha}\sim(\varphi_c-\varphi)^{-\alpha}$, with $\alpha\approx2.5$ for hard sphere glasses [\cite{Megen1994}] and assuming a range of values for weakly attractive gels.[\cite{Segre2001}] Likewise, power law divergence in the gel plateau modulus is a hallmark of attractive colloidal gels as a function of volume fraction, $G'_p\sim\delta_{\varphi}^{\nu_{\varphi}}$, and interaction strength, $G'_p\sim\delta_{U}^{\nu_{U}}$ where $\delta_{\varphi}=\varphi/\varphi_c -1$ is the normalized distance from the transition.[\cite{Trappe2000}] The exponents observed are not universal, but depend quite sensitively on the details of the interaction potential, both range, and magnitude, but typically $2\lesssim\nu_{\varphi}\lesssim4$ in weakly attractive systems[\cite{Prasad2003}]. Likewise, the critical volume fraction is a function of the interaction potential and vice-versa. The author is unaware of any work to date that has addressed potential divergence in relaxation time or modulus as a function of stress, although there has been elegant work on stress-delayed solidification [\cite{Ovarlez2008}], delayed yielding [\cite{Lindstrom2012,Sprakel2011}] as well as viscosity bifurcation [\cite{DaCruz2002,Coussot2002,Fielding2009,Mansard2011}] that relate well to this topic, with the latter two examples in essence traversing the fluid-solid transition in reverse.

At present it is not clear whether or how the experimentally observed critical behavior of relaxation time and shear modulus as a function of $\varphi$ and $U$ described above can be strictly mapped onto the critical behavior associated with gelation by bond percolation, or extent of reaction, $p$.[\cite{Prasad2003}] The same holds true for results obtained by the application by mode coupling theory to dense systems with attractive interactions, attractive glasses.[\cite{Bergenholtz1999,bergenholtz1999gel,Dawson2001,Puertas2005}] $\varphi$ should be related to the extent of reaction or probability of bond formation, $p$ but the exact form of the relationship remains unclear. A direct mapping between concentration and extent of reaction is present in the work of Aoki \textit{et al}. with $\epsilon=p/p_c-1=c/c_g-1$ where $c_g$ is the gel concentration. They recovered $k=1.5$ for the divergence of zero shear viscosity in a physical gel of semicrystalline PVC in solution [\cite{Aoki_Li1997b}] and $n=0.75$ for the relaxation exponent.[\cite{Aoki_Li1997a}] It would seem that the most natural mapping experimentally is between the extent of reaction and lab-time, where over a small interval $\Delta p$, it is presumed that $p\sim t$, and this has been verified experimentally.[\cite{venkataraman1989critical}] This provides a strong motivation for examining fluid-solid transitions in colloidal systems as a function of time. It is worth noting however that experimental data on volume fraction dependence of the elastic modulus for silica colloids in decalin shows $G'\sim(\varphi/\varphi_g-1)^{\nu_{\varphi}}$, with $\nu_{\varphi}=1.7$ by [\cite{ramakrishnan2006microstructure}] and $\nu_{\varphi}=2.0$ by [\cite{rueb1997viscoelastic}], both similar to the value $z=2$ derived here.

The similarities between chemical gelation in polymer systems and our colloidal gel as expressed through the critical exponents are certainly notable. On the basis of prior reports [\cite{Guo_Leheny2011,Eberle2012}], we expect the fractal dimension of the gels considered here to be $d_f\approx 2$. $d_f$ can be related to the relaxation exponent $n$ as suggested by Muthukumar for the case of polymer fractals with screened excluded volume [\cite{Muthukumar1989}], Equation \ref{eq:n_df}, where $d$ is the dimension of space. On this basis, the observed critical relaxation exponent of $n_c=0.5$ is consistent with a critical fractal dimension $d_f=2$, but it is not necessarily the case that the fractal dimension of the critical state be identical to the long time value, and so the agreement between $n_c$ and the expected fractal dimension of the gel at long time may be entirely coincidental. Additionally, it is not clear that Equation \ref{eq:n_df} is valid for colloidal gels, though it has been applied with some success.[\cite{Eberle2012}]

\begin{equation}
n=\frac{d(d+2-2d_f)}{2(d+2-d_f)}
\label{eq:n_df}
\end{equation}

Our data on these thermoreversible silica gels provides the first direct measurement of the critical dynamic exponent $\kappa$ for a colloidal gel. In combination with the critical relaxation exponent $n_c$ we can surmise on the basis of percolation theory that $k=z=2$, describing the divergence of viscosity and equilibrium modulus in the vicinity of $t_g$. While $\kappa$ and the proportionality between $(\partial \log{G'}/\partial t)$ and $(\partial \log{G''}/\partial t)$ are remarkably similar to those observed in polymer gels, the lack of effective microscopic models to provide specific predictions for the critical scaling exponents for gelation in particulate systems leaves us without a more appropriate point of reference. Direct measurements of the divergence of zero-shear viscosity and equilibrium modulus as a function of time are generally infeasible in particulate gels, making $k$ and $z$ effectively inaccessible. Exceptions to this in the context of volume fraction dependence are the use of scaling laws to estimate the $G'_p$ at arbitrarily small $\varphi$ [\cite{Trappe2000}], and the use of dynamic scattering to resolve the divergence of the alpha relaxation time. $n_c$ can be estimated quite readily by rheological or dynamic scattering experiments in near critical gels. Time-resolved data can provide access to $\kappa$ as demonstrated here, which provides the means to fully describe the critical behavior of colloidal gels as a function of reaction extent, or time.

By contrast, it is apparent that the colloidal glass considered here does not traverse a critical state as a function of time after cessation of flow. The solidification time or crossover time in this system displays a significant frequency dependence which can be viewed as a reflection of the length scale dependence in the relaxation behavior of the system. The lack of intersection in the time evolution of the loss tangent means there is no inversion in the frequency dependence of $\tan(\delta)$ and that the system remains effectively a liquid, despite the dramatic reduction of mobility and the emergence of significant elasticity. This resonates with the conventional notion of a glass as displaying an arbitrarily large viscosity ($\eta>10^{13}$ Pa.s) with a liquid-like structure factor.[\cite{Ediger1996}]

\section{Conclusions}

We have examined the evolution of dynamic moduli in a thermoreversible colloidal gel and a soft repulsive colloidal glass undergoing dynamical arrest after cessation of a shear flow. The nature of this transition is qualitatively different in the gel compared to the glass, with the former passing through a critical intermediate state en route to a well-defined viscoelastic solid while the latter does not display the inversion in frequency dependence of loss tangent that would be expected for a transition to a ``true'' viscoelastic solid. The gel time $t_g$ is an exponential function of temperature and serves as a single parameter to capture the kinetics such that $G(t)$ can be scaled onto a single curve for different temperatures and the growth rate of the critical gel modulus also falls on a master curve. We identify a critical intermediate state that displays a critical relaxation exponent $n_c=0.5$ and dynamic exponent $\kappa=0.25$. Despite dramatically differing gel times, the critical gel strength $S_c\approx0.33$ is independent of temperature over the small range of temperatures considered. Our work demonstrates that measurements of critical exponents in colloidal gels can be advanced by careful time-resolved experiments and that the colloidal gels and glasses exhibit non-trivial distinctions in the slowing of their dynamics on cessation of flow. Future efforts examining dynamical arrest along the colloidal gel to attractive glass line as well as potential critical scaling of static and dynamic properties as a function of shear stress are expected to be fruitful.


\begin{acknowledgments}
The authors gratefully acknowledge NSF support - C.O. through CBET-0828905 and CBET-1066904, and S.R. through HRD-1238524 and CBET-1336166.
\end{acknowledgments}

\bibliography{viscoelasticity_gelation}
\end{document}